# Quasi-Probability Distribution Functions for Optical-Polarization

Ravi S. Singh[1], Sunil P. Singh[2], Gyaneshwar K. Gupta[1]

[1]Department of Physics, DDU Gorakhpur University, Gorakhpur-273009 (U.P.) India
[2]Department of Physics, K. N. I. P. S. S., Sultanpur (U. P.) India

**Abstract**

Cahill-Glauber C(s)-correspondence is employed to construct Quasi-Probability Distribution Functions (QPDFs) for optical-polarization in phase space following equivalent description of polarization in Classical Optics. The proposed scheme provides pragmatic insights as compared to obscure SU (2) quasi-distributions on Poincare sphere. QPDF (Wigner function) of bi-modal quadrature coherent states is evaluated and numerically investigated to demonstrate the application.

## 1. Introduction

Polarization in Optics is an age-old concept discovered by Danish Mathematician E. Bartholinus and interpreted by Dutch Physicist C. Huygens while investigating birefringence in Quartz crystal[1]. Classically, optical-polarization is defined as the temporal-evolution of tip of electric field vector (light vector), which, in general, traverses an ellipse of non-random eccentricity and orientation[2]. On varying 'ratio of amplitudes' and 'difference in phases' (characteristic parameters for optical-polarization[3]) of the components of electric field vector, at any spatial point, along transverse orthogonal bases of description, the polarization-ellipse degenerates into various states of polarization such as linear and circular polarizations. Several techniques such as Stokes-parameters, Jones matrix, Mueller matrix and Coherency matrix[4] have been devised to quantify optical-polarization. Due to feasibility of the experimental measurements and its generalization through operatic-modification in quantum-domain[5], Stokes-parameters are still affording operational-characterization for polarization of light in quantum optics.

Theory of optical-polarization is generally dealt with two extreme cases: Perfect polarized light and Unpolarized light. In between these extremities there exist infinitely many partial optical-polarized states. Until 1970's confusions regarding poignant definition of unpolarized light prevails. In 1971 Prakash and Chandra[6], and independently Agarwal[7], discovered the structure of density operator for unpolarized light (natural light) providing its complete statistical description. Lehner et al.[8] and others[9-10] re-visited unpolarized light offering some new insights. The authors in Ref.[6] demonstrated that the characterization of optical-polarization by Stokes parameters is inadequate for they encompass only linear interaction between optical field-modes and, thus, rendering only second-order effects in optical-polarization. Insufficiency of these parameters is critically observed in recent studies[11-12].

Mehta and Sharma[13] defined perfect polarized state of a plane monochromatic light by demanding vanishing amplitude in at least one orthogonal polarized mode after passing through a compensator (phase-shifter) followed by a rotator (polarizer). Such a polarized optical field is truly a single-mode field (having no signal in orthogonal mode) described by single random complex amplitude (CA). A criterion[14] for perfect optical-polarization state with the help of 'index of polarization' (IOP) is set up by the ratio of CAs in two transverse orthogonal bases-modes of which non-random values provide 'ratio of real amplitudes' and 'difference in phases', characteristic parameters for polarized light[3].

Stokes operators characterizing optical-polarization, offering analogous mathematical appearance as Jordan- Schwinger spin- angular momentum[15], obeys SU (2) Lie algebra. Tracing these sorts of similarities Luis[16] deduced the SU(2) Q-quasi distribution function over Poincare sphere for describing optical-polarization and defining degree of polarization in Quantum Optics by introducing abstract notion of distance from the Q-QPDF of unpolarized light to that of quantum states in question. Karassiov[9,17] has critically emphasized that Stokes operator describing optical-polarization states found distinct kind of Hilbert space for its operation compared to that of spin-angular momentum. The preceded discussion demands afresh description of optical-polarization.

The aim of present paper is to derive QPDF, especially Wigner function, with the aid of C(s) correspondence rule discovered by Cahill and Glauber in 1969. The presentation is organized as follows: in Section 2 basics of IOP are reviewed. Section 3 deals with Transiting operator which governs the transition from Hilbert space to Phase space. QPDF (Wigner distribution function) of bi-modal quadrature coherent states of light is evaluated in Section 4.

## 2. Index of Optical-Polarization

A plane monochromatic unpolarized optical field propagating along z-direction can, in general, be described by vector potential, $\mathcal{A}$ (analytical signal),

$$\mathcal{A} = (\hat{\mathbf{e}}_x \underline{A}_x + \hat{\mathbf{e}}_y \underline{A}_y) e^{-i\psi}, \underline{A}_{x,y} = A_{0x,0y} e^{i\phi_{x,y}}, \tag{1}$$

where $\psi = \omega t - kz$, $\underline{A}_{x,y}$ are classical CAs, $A_{0x,0y}$ are real amplitudes possessing, in general, random spatio-temporal variation with angular frequency, $\omega$, $\phi_{x,y}$ (phase parameters) may take equally probable random values in the interval $0 \leq \phi_{x,y} < 2\pi$ in linear polarization-basis $(\hat{\mathbf{e}}_x, \hat{\mathbf{e}}_y)$, $\mathbf{k}$ ($= k\hat{\mathbf{e}}_z$) is propagation vector of magnitude k, and $\hat{\mathbf{e}}_{x,y,z}$ are unit vectors along respective x-, y-, z-axes forming right handed triad.

Light may be said to be polarized only when the ratio of CAs must keep non-random values, i.e., $\underline{A}_y/\underline{A}_x = p$, a non-random complex parameter defining 'index of polarization' (IOP)[14]. Clearly, polarized optical field is a single mode field as only one random CA suffices for its complete statistical description (other orthogonal CA is specified by p). If one introduces new parameters $A_0$ (real random amplitude), $\chi_0$ (polar angle), $\Delta_0$ (azimuth angle), $\phi$ (random phase) on a Poincare sphere, satisfying inequalities $0 \leq A_0$, $0 \leq \chi_0 \leq \pi$, $-\pi < \Delta_0 \leq \pi$, $0 \leq \phi < 2\pi$, respectively, and preserving transforming equations in terms of old parameters, $A_0 = (A_{0x}^2 + A_{0y}^2)^{1/2}$; $\chi_0 = 2\tan^{-1} A_{0y}/A_{0x})$ and $\Delta_0 = \phi_y - \phi_x$; $\phi = (\phi_x + \phi_y)/2$, the analytic signal, $\mathcal{A}$, Eq. (1), finds an instructive compact form

$$\boldsymbol{\mathcal{A}} = \hat{\boldsymbol{\varepsilon}}_0 \mathcal{A}; \mathcal{A} = \underline{A} e^{-i\psi}; \underline{A} = A_0 e^{i\phi},$$

$$\hat{\boldsymbol{\varepsilon}}_0 = \hat{\mathbf{e}}_x \cos\tfrac{\chi_0}{2} e^{-\Delta_0/2} + \hat{\mathbf{e}}_y \sin\tfrac{\chi_0}{2} e^{-\Delta_0/2}. \qquad (2)$$

The Eq. (2) may be interpreted as the single mode polarized optical field, statistically described by single CA, $\underline{A}$ and polarized in the fixed direction, $\hat{\boldsymbol{\varepsilon}}_0$ determined by non-random angle parameters $\chi_0$ and $\Delta_0$ in the Poincare sphere specifying the mode, $(\hat{\boldsymbol{\varepsilon}}_0, \mathbf{k})$. The complex vector $\hat{\boldsymbol{\varepsilon}}_0$ is a unit vector ($\hat{\boldsymbol{\varepsilon}}_0^* \cdot \hat{\boldsymbol{\varepsilon}}_0 = 1$) giving expression of IOP, $p = \tan\tfrac{\chi_0}{2} e^{i\Delta_0}$. Obviously, the state of polarization of light is specified by the non-random values of p, which, in turn, is fixed by non-random values of $\chi_0$ and $\Delta_0$ defining a point $(\hat{\boldsymbol{\varepsilon}})$, in the unit Poincare sphere, similar to Stokes parameters. All typical parameters in ellipsometry such as major axis, minor axis and orientation angles of the polarization-ellipse can be determined if p of optical field and one CA are specified. In elliptic-polarization basis $(\hat{\boldsymbol{\varepsilon}}, \hat{\boldsymbol{\varepsilon}}_\perp)^{18}$ such polarized light (Eq. 2) retain IOP, $p_{(\hat{\boldsymbol{\varepsilon}},\hat{\boldsymbol{\varepsilon}}_\perp)}$, another non-random parameter showing vivid dependence on $\hat{\boldsymbol{\varepsilon}}_0$,

$$p_{(\hat{\boldsymbol{\varepsilon}},\hat{\boldsymbol{\varepsilon}}_\perp)} = \underline{A}_{\hat{\boldsymbol{\varepsilon}}_\perp}/\underline{A}_{\hat{\boldsymbol{\varepsilon}}} = \frac{\hat{\boldsymbol{\varepsilon}}_\perp^* \cdot \hat{\boldsymbol{\varepsilon}}_0}{\hat{\boldsymbol{\varepsilon}}^* \cdot \hat{\boldsymbol{\varepsilon}}_0}, \qquad (3)$$

where $\hat{\boldsymbol{\varepsilon}}_\perp$ is orthogonal complex unit vector to $\hat{\boldsymbol{\varepsilon}}$, i. e. $\hat{\boldsymbol{\varepsilon}}^* \cdot \hat{\boldsymbol{\varepsilon}} = \hat{\boldsymbol{\varepsilon}}_\perp^* \cdot \hat{\boldsymbol{\varepsilon}}_\perp = 1$, $\hat{\boldsymbol{\varepsilon}}_\perp^* \cdot \hat{\boldsymbol{\varepsilon}} = 0$.

In Quantum Optics the optical field, Eq. (1) is described by operatic-version of vector potential operator,

$$\widehat{\mathcal{A}} = \left(\tfrac{2\pi}{\omega V}\right)^{1/2} [(\hat{\mathbf{e}}_x \hat{a}_x + \hat{\mathbf{e}}_y \hat{a}_y) e^{-i\psi} + \text{h.c.}],$$

$$= \left(\tfrac{2\pi}{\omega V}\right)^{1/2} [(\hat{\boldsymbol{\varepsilon}} \hat{a}_{\hat{\boldsymbol{\varepsilon}}} + \hat{\boldsymbol{\varepsilon}}_\perp \hat{a}_{\hat{\boldsymbol{\varepsilon}}_\perp}) e^{-i\psi} + \text{h.c.}], \qquad (4)$$

in linear-polarization basis $(\hat{\mathbf{e}}_x, \hat{\mathbf{e}}_y)$ or in elliptic-polarization basis $(\hat{\boldsymbol{\varepsilon}}, \hat{\boldsymbol{\varepsilon}}_\perp)$, respectively, where $\omega$ is angular frequency of the optical field and V is the quantization volume of the cavity, h.c. stands for Hermitian conjugate. Using orthonormal properties of $\hat{\boldsymbol{\varepsilon}}$ ($= \varepsilon_x \hat{\mathbf{e}}_x +$ $ε_y\hat{e}_y$) and $\hat{ε}_⊥(= ε_{⊥x}\hat{e}_x + ε_{⊥y}\hat{e}_y)$ the annihilation operators $\hat{a}_{\hat{ε}}$ ($\hat{a}_{\hat{ε}_⊥}$) are related with those in linear-polarization basis $(\hat{e}_x, \hat{e}_y)$ by,

$$\hat{a}_{\hat{ε}} = ε_x^* \hat{a}_x + ε_y^* \hat{a}_y, \quad \hat{a}_{\hat{ε}_⊥} = ε_{⊥x}^* \hat{a}_x + ε_{⊥y}^* \hat{a}_y, \tag{5}$$

satisfying usual Bosonic-commutation relations. By demanding non-random values of all multiple powers of positive integer, say, n of ratio of CAs except n = 1(Eq.3) the concept of Higher-order optical-polarization has been introduced[19]. Moreover, the Glauber coherence functions[20] describe correlation properties of optical-field at any spatio-temporal point, providing interference effects, are given by,

$$Γ^{(m_x,m_y,n_x,n_y)} = \text{Tr}[ρ(0)\hat{A}_x^{(-)m_x}\hat{A}_y^{(-)m_y}\hat{A}_x^{(+)n_x}\hat{A}_y^{(+)n_y}] \tag{6}$$

where $ρ(0)$ is density operator describing dynamical state of optical field. Setting the condition on quantized complex amplitudes in Eq.(4),

$$\hat{a}_y(t)ρ(0) = p\,\hat{a}_x(t)ρ(0), \tag{7}$$

where p is IOP for polarized light, one obtains after inserting Eq.(7) into (6),

$$Γ^{(m_x,m_y,n_x,n_y)} = p^{*m_y}p^{n_y}Γ^{(m_x+m_y,0,n_x+n_y,0)} \tag{8}$$

which describes correlation properties of single-mode optical-polarization state. Clearly, Glauber coherence function, Eq.(8) is determined by p (IOP) and one of quantized complex amplitudes $\hat{a}_x(t)$. Notably, since, Eq.(7) gives the coherence function for polarized light, Eq.(7) may be regarded as quantum analogue of classical criterion $\underline{A}_y = p\,\underline{A}_x$ for optical-polarized field. Conclusively, one may infer that the concept of index of optical-polarization reduces bimodal description of optical-polarization into mono-modal if indices are provided.

**3. Transiting Operator**

In Classical Optics, optical-polarization is described by superposition of two transverse harmonic oscillators of same frequencies endowed with non-random values of 'ratio of real amplitudes' and 'difference in phases'. The parametric equations for CAs are

$$α_x(\underline{A}_x) = A_0 \cos(χ_0/2) \exp(iφ_x) \text{ and } α_y(\underline{A}_y) = A_0 \sin(χ_0/2) \exp(iφ_y), \tag{9}$$

where $A_0^2$ ($= |\alpha_x|^2 + |\alpha_y|^2$) is intensity of light. Obviously the IOP,

$$p = \alpha_y / \alpha_x = \tan\frac{\chi_0}{2} e^{i\Delta_0}, \qquad (10)$$

In quantum optics the roles of classical CAs ($\alpha_{x,y}$) of optical field is taken by Bosonic annihilation operators ($\hat{a}_{x,y}$) for quantized complex amplitudes of each transverse quantum oscillators of same frequencies which bear the relationship,

$$\hat{a}_y \rho = p\, \hat{a}_x \rho \qquad (11)$$

to describe perfect polarized light (cf. Eq.7).

Cahill-Glauber[21] introduced C(s) correspondence between bounded operators and its weight functions. Accordingly, the dynamic state of a single harmonic oscillator of classical complex amplitude, $\alpha$ is described in quantum theory by its density operator, $\hat{\rho}$ in Hilbert space of which description in phase space is governed by QPDF $W(\alpha, s)$ such that

$$W(\alpha, s) = \text{Tr}[\hat{\rho}\, \hat{t}(\alpha, s)] \qquad (12)$$

where $\hat{t}(\alpha, s)$ are Kernel-operators, defined as complex Fourier transform of the s-ordered displacement operators $D(\alpha, s)$, given by,

$$\hat{t}(\alpha, s) = \int \hat{D}(\xi, s)\, \exp(\alpha\xi^* - \alpha^*\xi)\, \pi^{-1}\, d^2\xi. \qquad (13)$$

Here $\hat{D}(\xi, s)$ bears the relation with ordinary unitary displacement operator, $\hat{D}(\xi)$,

$$\hat{D}(\xi, s) = \hat{D}(\xi) \exp(s\, |\xi|^2 / 2), \qquad (14)$$

Simple algebraic manipulation provides expression for, $\hat{t}(\alpha, s)$ as

$$\hat{t}(\alpha, s) = \left(\frac{2}{1-s}\right) \hat{D}(\alpha) \left(\frac{s+1}{s-1}\right)^{\hat{a}^\dagger \hat{a}} \hat{D}^\dagger(\alpha), \qquad (15)$$

or, in equivalent separable form as

$$\hat{t}(\alpha, s) = \left(\frac{2}{1-s}\right) \exp\left(\frac{-2\,|\alpha|^2}{1-s}\right) \exp\left(\frac{2\alpha\,\hat{a}^\dagger}{1-s}\right) :\exp\left(\frac{-2\hat{a}^\dagger \hat{a}}{1-s}\right): \exp\left(\frac{2\alpha^*\,\hat{a}}{1-s}\right), \qquad (16)$$

where : : represents normal ordering, (see Eqs.(I.6.23) and (I.6.33) in Ref.21). The phase space description of the two independent transverse harmonic oscillators may, correspondingly, be transited by operator $\hat{T}(\alpha_x, \alpha_y, s)$ defined to be

$$\hat{T}(\alpha_x, \alpha_y, s) = \hat{t}(\alpha_x, s)\hat{t}(\alpha_y, s), \tag{17a}$$

insertion of Eq. (16) into Eq.(17) for each oscillators one obtains expression for transiting operator,

$$\hat{T}(\alpha_x, \alpha_y, s) = \left(\frac{2}{1-s}\right)^2 \exp\left(\frac{-2(|\alpha_x|^2 + |\alpha_y|^2)}{1-s}\right) \exp\left(\frac{2(\alpha_x \hat{a}_x^\dagger + \alpha_y \hat{a}_y^\dagger)}{1-s}\right)$$

$$:\exp\left(\frac{-2(\hat{n}_x + \hat{n}_y)}{1-s}\right): \exp\left(\frac{2(\alpha_x^* \hat{a}_x + \alpha_y^* \hat{a}_y)}{1-s}\right), \tag{17b}$$

Eq.(17b) may be regarded as Transiting Operator (TO) which governs projection of density operator ρ for two independent transverse oscillators from Hilbert space to Quantum Phase space.

## 4. Phase Space Description of Bi-modal Quadrature Coherent States

Since polarization of optical field is characterized by IOP, p, Eq.(10) in Classical Optics, the operator, Eq.(17b) which facilitates the mapping provided by Eq.(12) yields the form,

$$\hat{T}(\alpha_x, p\alpha_x, s) = \left(\frac{2}{1-s}\right)^2 \exp\left(\frac{-2(1+|p|^2)|\alpha_x|^2}{1-s}\right) \exp\left(\frac{2\alpha_x(\hat{a}_x^\dagger + p\hat{a}_y^\dagger)}{1-s}\right)$$

$$:\exp\left(\frac{-2(\hat{n}_x + \hat{n}_y)}{1-s}\right): \exp\left(\frac{2\alpha_x^*(\hat{a}_x + p^*\hat{a}_y)}{1-s}\right). \tag{18}$$

Let us derive QPDF utilizing Eq.(18) by considering light in the quadrature coherent state described by density operator $\hat{\rho}$,

$$\hat{\rho} = |\beta, \gamma\rangle\langle\beta, \gamma| \tag{19}$$

where quadrature oscillators are described by classical amplitude β and γ. The application of Eq.(7) gives IOP, q as

$$q = \gamma/\beta = \tan(\theta/2) e^{i\delta}, \tag{20}$$

after parametrizing β and γ as

$$\beta = B_0 \cos(\theta/2) \exp(-i\delta/2 + i\varphi); \quad \gamma = B_0 \sin(\theta/2) \exp(i\delta/2 + i\varphi), \tag{21}$$

insertion of Eq.(19) along with (20) and (10) for indices of polarization in (12) yields, after use of (18) and simple manipulation, s-parametrized QPDF as

$$W(|\alpha_x|, |\beta|, s) = \langle \beta\gamma|\hat{T}(\alpha_x, p\alpha_x, s)|\beta\gamma\rangle$$

$$= \left(\frac{2}{1-s}\right)^2 \exp\left(\frac{-2|\alpha_x - \beta|^2}{1-s}\right) \exp\left[\frac{2}{1-s}\left\{-\left(|\alpha_x|\tan\left(\frac{\chi_0}{2}\right) + |\beta|\tan\left(\frac{\theta}{2}\right)\right)^2 + \tan\left(\frac{\chi_0}{2}\right)\tan\left(\frac{\theta}{2}\right)\left(\sqrt{\alpha_x\beta^*}e^{\frac{i(\Delta_0-\delta)}{2}} + c.c.\right)^2\right\}\right] \quad (22)$$

Eq. (22) is Gaussian and may be shown to satisfy normalization condition for being QPDF. Eq. (22) may be integrated over intensity (which decides the radius of Poincare sphere) yielding QPDF on Poincare sphere. Similar calculation for quadrature coherent state of equal amplitudes is obtained by Klimov et.al.[22].

Considering s = 0 in Eq.(22) Wigner distribution function can be obtained. Winger distribution is plotted against the variation of phases (arg α) at the fixed value of amplitude of optical field, which has Gaussian shape (Figs.1a, b). It is also shown how Wigner distribution changes with the amplitude of light on some fixed value of phases (Fig.2 c, d)

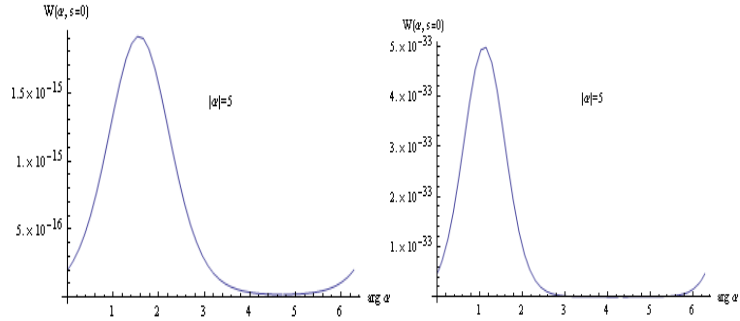

**Fig.1** Wigner distribution (s=0) exhibiting dependences on the various phases, arg α with fixed value of, (a) |α|=5; p=q= 0.0049(1+i); β=2 $e^{i\pi/2}$ (b) |α|=5; p=q= $2^{-1/2}$(1+i) ; β=$20^{-1/2}$ $e^{i\tan^{-1}(2)}$

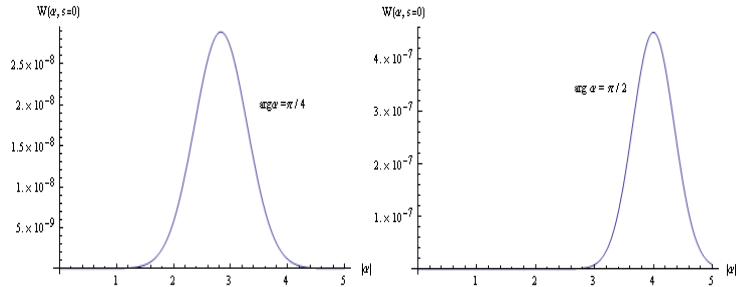

**Fig.2** Wigner distribution (s=0) exhibiting dependences on the variation of |α| with fixed value of, (c) arg α = π/4; $p_c=p_q$= 0.0049(1+i); β=2 $e^{i\pi/2}$ (d) arg α = π/2; $p_c=p_q$= $2^{-1/2}$(1+i) ; β=$20^{-1/2}$ $e^{i\tan^{-1}(2)}$

**Acknowledgement:** The authors are highly obliged for critical yet inspirational suggestions imparted by Prof. H. Prakash and R. Prakash.

Additional authors information:  #yesora27@gmail.com,  *gyankg@gmail.com

## References

[1]   Brosseau C., Polarization and Coherence Optics: Historical Perspective, Status and Future Directions, Progress in Optics Vol. 54, Ed. E. Wolf, Elsevier Publication, Great Britain, 149-153 (2009).
[2]   Born M. and Wolf E., [Principles of Optics] Cambridge University Press, Cambridge (1999).
[3]   Peres A., [Quantum Theory: Concept and Methods], Kluwer Academic Publishers, London, U. K. (2002).
[4]   Collett E., [Polarized Light: Fundamentals and Applications] Marcel Dekkers Inc., NewYork, (1993).
[5]   Collett E., "Stokes Parameters for Quantum systems", Am. J. Phys. 38, 563 (1970).
[6]   Prakash H. and Chandra N., "Density Operator of Unpolarized Radiation", Phys. Rev. A 4, 796 (1971).
[7]   Agarwal G. S., "On the State of Unpolarized Radiation", Lett. Nuovo Cim. 1, 53 (1971).
[8]   Lehner J., Leonhardt U. and Paul H., "Unpolarized light: Classical and quantum states", Phys. Rev. A 53, 2727 (1996).
[9]   Karassiov V. P., "Polarization Structure of Quadrature Light Fields: a new insight.I. General Outlook", J. Phys. A: Math. Gen. 26, 4345 (1993).
[10]  Soderholm J.and Trifonov A., "Unpolarized light in quantum optics", Opt. Spectrosc. 91, 532 (2001).
[11]  Bjork G., Soderholm J., Sanchez-Soto L. L., Klimov A. B., Ghiu I., Marianand P. and Marian T. A., "Quantum Degrees of Polarization" Opt. Commn. 283, 4400 (2010).
[12]  Klimov A. B., Bjork G., Soderholm J., Madsen L. S., Lassen M., Andersen U. L., Heersink J., Dong R., Marquardt Ch., Leuchs G. and Sanchez-Soto L. L., "Assessing the Polarization of a Quantum Field from Stokes Fluctuations" Phys. Rev. Lett., 105, 153602 (2010).
[13]  Mehta C. L. and Sharma M. K., "Diagonal Coherent State Representation for Polarized Light", Phys. Rev. D 10, 2396 (1974).
[14]  Prakash H. and Singh Ravi S., "Operator for Optical Polarization", J. Phys. Soc. Japan, 69, 284 (2000).
[15]  Schwinger J., "Unitary Operator Bases", Proc. Natl Acad. Sci. USA 46, 570 (1960).
[16]  Luis A., "Quantum Polarization Distribution via Marginals of Quadrature Distributions", Phys. Rev. A 71, 053801 (2005).
[17]  Karassiov V. P., "Polarization Squeezing and new States of Light in Qunatum Optics", Phys. Lett. A, 190, 387 (1994).
[18]  Mandel L. and Wolf E., [Optical Coherence and Quantum Optics] Cambridge University Press, Cambridge, 468-471 (1995).
[19]  Singh Ravi S. and Prakash H., arXiv:1103.4243v1.
[20]  Glauber R. J., "The Quantum Theory of Optical Coherence", Phys. Rev. 130, 2529; ibid "Coherent and Incoherent States of Radiation Field", 131, 2766 (1963).
[21]  Cahill K. E. and Glauber R. J., "Ordered Expansions in Boson Amplitude Operators; Density Operators and Quasiprobability Distribution", Phys. Rev., 177, 1857, 1882 (1969).
[22]  Klimov A. B. et. al., "Quantum Phase-Space Description of Light Polarization", Opt. Commun. 258, 210 (2006).